\documentstyle[12pt]{article}
\newcommand {\beq} {\begin{equation}}
\newcommand {\eeq} {\end{equation}}

\begin{document}
\input epsf
\noindent
{\Large\bf Stress-Energy Must be Singular on
the Misner Space Horizon even for Automorphic Fields}    
\vskip 1 cm

\noindent
Claes R Cramer\footnote{email:crc101@york.ac.uk} 
and Bernard S. Kay\footnote{email:bsk2@york.ac.uk}\\
\noindent
Department of Mathematics, University of York, Heslington, York YO1 5DD,
UK
\vskip 1 cm
\noindent
{\bf Abstract.} We use the image sum method to reproduce Sushkov's
result that for a massless automorphic field on the initial globally
hyperbolic region $IGH$ of Misner space, one can always find a special
value of the automorphic parameter $\alpha$ such that the renormalized 
expectation value $\langle\alpha|T_{ab}|\alpha\rangle$ in the {\it 
Sushkov state} ``$\langle\alpha|\cdot|\alpha\rangle$'' (i.e. the
automorphic generalization of the Hiscock-Konkowski state) vanishes.
However, we shall prove by elementary methods that  the conclusions of a
recent general theorem of Kay-Radzikowski-Wald apply in this case.  I.e.
 for any value of $\alpha$ and any neighbourhood $N$ of any point $b$ on
the chronology horizon  there exists at least one pair of non-null
related points  $(x,x') \in (N\cap IGH)\times (N\cap IGH)$ such that the
renormalized two-point function of an automorphic  field $G^\alpha_{\rm
ren}(x,x')$ in the Sushkov state is singular.  In consequence
$\langle\alpha|T_{ab}|\alpha\rangle$ (as well as other  renormalized
expectation values such as  $\langle\alpha|\phi^2|\alpha\rangle$) is
necessarily singular {\it on} the chronology horizon.  We point out that
a similar situation (i.e. singularity {\it on} the chronology horizon)
holds for states on Gott space and Grant space.
\vskip 1 cm 
\noindent 
There has recently been interest in quantum field theory on spacetimes
with a chronology horizon. (See \cite{Thorne92,Hawking92,Kay96} and 
references therein.) The reason for this interest is that these
spacetimes (or rather those for which the chronology horizon is
compactly generated \cite{Hawking92}) are
models for spacetimes in which time machines get manufactured. Hawking
\cite{Hawking92} has proposed the  {\it Chronology Protection
Conjecture} according to which the laws of physics (and in particular
quantum effects) prevent such  spacetimes from being physically
realizable.  In  fact, there were  arguments due to Hawking and others
(see e.g. references in  \cite{Kay96,Hawking92}) that, at least in the
case of simple model quantum field theories on such spacetimes
\cite{Hawking92} and for a wide range of states on the initial globally
hyperbolic region, the (expectation value of the renormalized)
stress-energy tensor would diverge as one approached the chronology
horizon.  We remark on the one hand that this evidence was of a
heuristic nature and it is now known not to be true in general that the
stress-energy tensor really does  diverge as one approaches the
chronology horizon for ``all'' initial states.  In fact, recently
Krasnikov \cite{Krasnikov95}, in the context of a scalar field on 
two-dimensional Misner space, and Sushkov in the context of {\it
automorphic} fields on four-dimensional Misner space have given examples
of (Hadamard) states for which the stress-energy {\it vanishes} on the
initial globally hyperbolic region.  (Additionally, Boulware
\cite{Boulware92}, in the case of Gott space, and Tanaka and Hiscock
\cite{Tanaka96}, in the case of Grant space, have shown that for a
sufficiently massive field, the stress-energy is bounded on the initial
globally hyperbolic region.   On the other hand, a slightly
different statement (Theorem 2 of \cite{Kay96}) has recently been
rigorously proven by Kay, Radzikowski and Wald: namely, that, for the 
model consisting of the real covariant Klein-Gordon equation on a
spacetime with compactly generated chronology horizon, for any Hadamard
state on the initial globally hyperbolic region, the expectation value
of the  renormalized stress-energy tensor (and also of other similarly
defined quantities such as the renormalized expectation value of
$\phi^2$) is necessarily singular {\it on} the chronology horizon. (See
before the statement of our theorem below for a more precise 
statement/discussion.)

We shall focus on Sushkov's model which concerns a generalization
\cite{Sushkov95}, which we shall call here the {\it Sushkov state},  of
the Hiscock-Konkowski state \cite{Hiscock82} for a massless automorphic
field on the initially globally hyperbolic region of (four-dimensional)
Misner space.  The Sushkov state $\langle\alpha|\cdot|\alpha\rangle$ is
labelled by a phase, $\alpha$, called the automorphic parameter and
Sushkov pointed out that there must exist a value of this automorphic
parameter for which the expectation  value
$\langle\alpha|T_{ab}|\alpha\rangle$ in the Sushkov state of the 
renormalized stress-energy tensor vanishes (and another value for which
the renormalized expectation value of $\phi^2$  vanishes). 

The Kay-Radzikowski-Wald theorem was not explicitly proved for the case
of automorphic fields.\footnote{The theorem also does not strictly apply
because the chronology horizon of four-dimensional Misner space is not
compactly generated.  However, one easily sees that the theorems of
\cite{Kay96} still hold in this case since the conclusions of
Proposition 2 of \cite{Kay96} still hold in the case of any spacetime
which arises as the product with a Riemannian manifold of dimension
$4-d$ of a spacetime with compactly generated chronology horizon with
dimension $d<4$.}  However, the extension to this case appears to be
straightforward.  In fact, one expects there to be no difficulty
\cite{Cramer96} in extending the  theorems of \cite{Kay96} to the case
of a complex scalar field in an external electromagnetic field, and the
case of an automorphic field may of course be regarded as a special case
of this. Applying this result to the situation discussed by
Sushkov, we thus face the rather surprising occurrence of an 
energy-momentum tensor which, while bounded (in fact zero!) {\it up to}
a (four-dimensional)  chronology horizon is nevertheless singular {\it
on} the chronology horizon with the conclusion that even in this case,
the Chronology Protection Conjecture does not, after all, appear to be
contradicted.  (We shall also explain at the end of this letter that the 
states proposed by Boulware and Tanaka-Hiscock for Gott space and Grant
space \cite{Grant92} will also have a stress-energy tensor which is 
singular on the chronology horizon in a slightly different sense, and for
slightly different reasons.)

Rather than elaborating on the details of the general extension of the
general theorem of Kay-Radzikowski-Wald, in the present letter we shall
provide an elementary proof that the conclusions of the
Kay-Radzikowski-Wald theorem apply to the specific situation discussed
by Sushkov, i.e. we shall show directly that for any value of the 
automorphic parameter $\alpha$, $\langle\alpha|T_{ab}|\alpha\rangle$ and
 $\langle\alpha|\phi^2|\alpha\rangle$ must necessarily be singular {\it
on} the chronology horizon of Misner space.  To accomplish this we
express the Sushkov two-point function for an  automorphic field by an
image-sum formula which generalizes that of Hiscock and  Konkowski
\cite{Hiscock82}.  

We remark that the proof of the theorem in \cite{Kay96} appeals to
powerful general  theorems of Duistermaat and H\"ormander on the
``Propagation of Singularities'' \cite{Hormander85} whereas, in the
specific situation discussed here, one may bypass any appeal to these
general theorems since one sees explicitly, by direct inspection of the
image sum formula, how this propagation of singularities occurs. 
In this way, we hope the discussion below provides a useful example of,
and could serve as an entr\'ee to, the general theorems of \cite{Kay96}.

A convenient mathematical description of (four dimensional) Misner 
space may be obtained by taking the region $\cal R$
\begin{eqnarray}
U\in (-\infty,0)\\
V\in (-\infty, \infty)
\end{eqnarray}
of Minkowski space with the metric  
\beq
ds^{2}=-dUdV+dY^{2}+dZ^{2}
\eeq
(where $U$ and $V$ are the double null coordinates $U=T-X$, $V=T+X$) and
identifying points related by a fixed Lorentz boost  $L$ (with rapidity
say $a$)  in the $(T,X)$ plane by making the identifications 
\beq
(U,V,Y,Z)\leftrightarrow (e^{-na}U,e^{an}V,Y,Z).\label{ident}
\eeq
By going to $(\tau,u)$ coordinates on $\cal R$,  where $\tau=-UV$,
$u=-\ln(-U)$ so that $L$ maps $(\tau,u)$ to $(\tau,u+a)$ one may see
that Misner space arises as the  product with two-dimensional Euclidean
space ($R^2$ with the metric $ds^2=dY^2+dZ^2$) of two-dimensional Misner
space ($R\times S$ with metric $ds^2=-\tau du^2 +d\tau du$, where $u$
ranges from $0$ to $a$ with $u=0$ and $u=a$ identified). The latter
consists (see Figure 1) of an initial globally hyperbolic region
($\tau<0$) $IGH_2$ (the quotient of region ``III'' of two-dimensional
Minkowski space by $L$) and a region of closed timelike curves
($\tau>0$) $CTC_2$ (the quotient of region ``I'' of  two-dimensional
Minkowski space by $L$) separated by the closed null geodesic ($\tau=0$)
$CNG$ (the quotient of the half null line separating region I from
region III). Restoring the product with two dimensional Euclidean space, 
$IGH_2$, $CTC$ and $CNG$ give rise respectively to regions which we
shall denote $IGH$, $CTC$ and $CH$ (i.e. ``chronology horizon'') of four
dimensional Misner space.\\\\                                                   
\def\epsfsize#1#2{0.5#1}                                                
\epsfclipon                                                             
\epsffile[70 310 775 725]{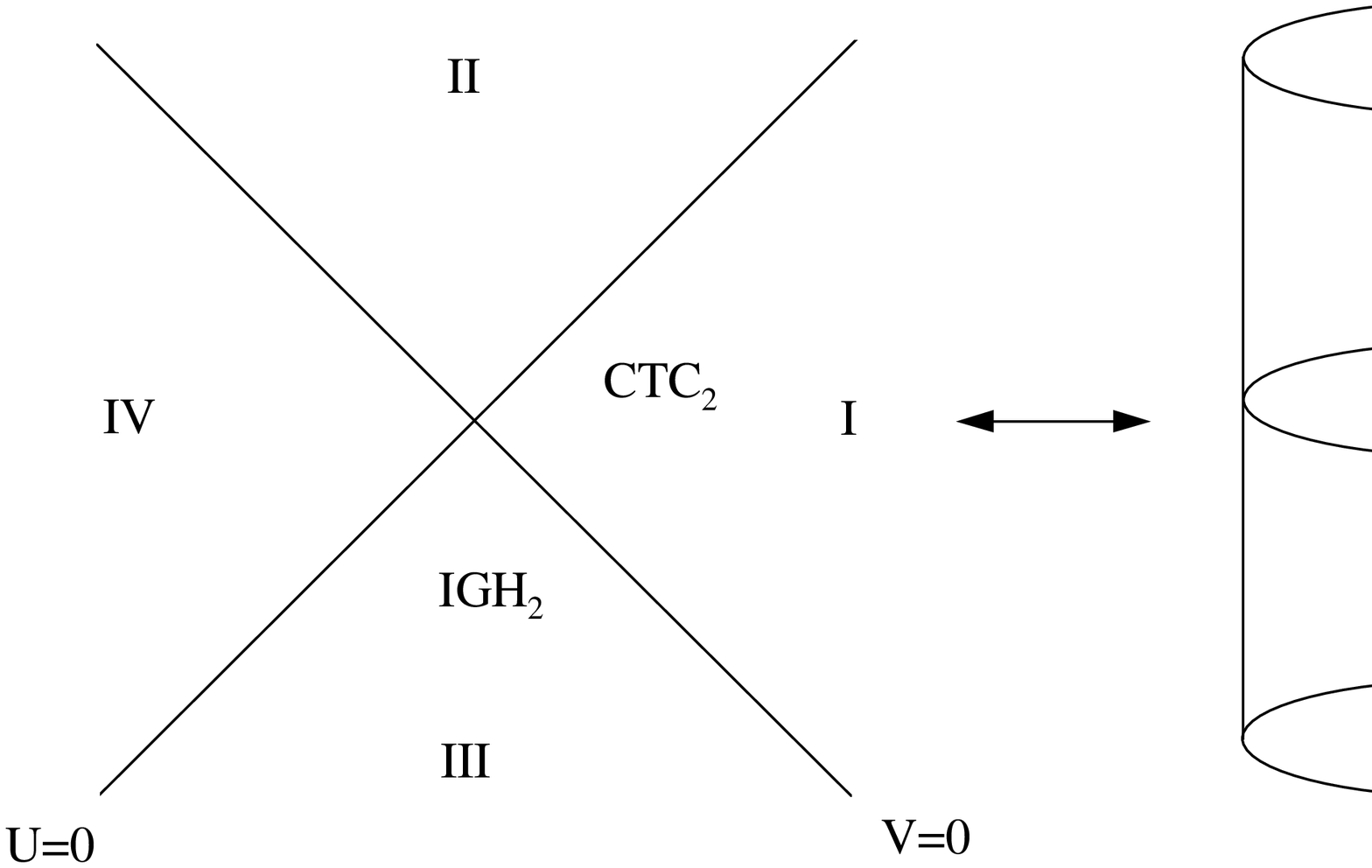}                                    
Fig1.\\\\                                                               
In terms of the representation of Misner space as the region $\cal R$ of
Minkowski space with the identifications (\ref{ident}) we may think of a
classical automorphic field as  a complex solution $\phi$ of the 
Klein-Gordon equation on  $\cal R$ satisfying the condition 
\beq
\phi(Lx)=e^{2\pi i\alpha}\phi(x)\label{auto}
\eeq
for all $x\in {\cal R}$ where $\alpha$ is a parameter lying in the interval
$[0,1)$.  (The case $\alpha=0$ corresponds to an ordinary complex
field, the case $\alpha={1\over 2}$ to a complex twisted field.)

For each $\alpha$, the Sushkov state
``$\langle\alpha|\cdot|\alpha\rangle$'' for a quantum complex
automorphic field on the initial globally hyperbolic region $IGH$ is
defined in \cite{Sushkov95} by a  mode-sum method.  It is possible to
show  \cite{Cramer96} that the two-point function $G^\alpha$ in this 
state arises as the image sum
\beq
G^\alpha(x,x'):=\langle\alpha|\phi(x)\phi^\dagger(x')|\alpha\rangle=
\sum_{n=-\infty}^{\infty}
e^{2\pi in\alpha}
G_{\rm M}(x,L^nx')
\label{image}
\eeq
where $G_{\rm M}(x,x')=\langle M|\phi(x)\phi^\dagger(x')|M\rangle$ is 
the two  point function in the Minkowski vacuum which, for non-null
separated pairs of points is equal to $(2\pi)^{-2}/\sigma(x,x')$ where
$\sigma$ is the square of the geodesic distance between $x$ and $x'$.  
In particular, in the case $\alpha=0$, the Sushkov two-point function
coincides with the Hiscock-Konkowski two-point function.  (We remark in
passing that this identity  of the Hiscock-Konkowski image sum with
Sushkov's mode sum settles in the affirmative the -- as far as we are
aware, hitherto open -- question as to whether the Hiscock-Konkowski
two-point function satisfies the positivity conditions required for it
to arise from a quantum state.)

Generalizing the point-splitting procedure of \cite{Hiscock82},  the
renormalized expectation value of the (conformally improved) 
stress-energy tensor in the Sushkov state is given by
\beq
\langle T_{ab}(y)\rangle=
\lim_{(x,x')\rightarrow (y,y)}
\left (\frac{2}{3}\partial_{a}\partial_{b'}
-\frac{1}{6}\eta_{ab}\eta^{cd'}\partial_{c}
\partial_{d'} -\frac{1}{3}\partial_{a}
\partial_{b} \right 
)G^\alpha_{\rm ren}(x,x') \label{unsplit}. 
\eeq
where
\beq
G^\alpha_{\rm ren}(x,x')=\frac{1}{(2\pi)^{2}}
\sum_{n=-\infty ,\ n\not =0}^{\infty}
\frac{e^{2\pi in\alpha}}
{\sigma(x,L^n x')}
\label {Gra}
\eeq
is the renormalized two-point function.

Using eq. (\ref{unsplit}) and (\ref{Gra}) we find that the renormalized 
stress-energy is given by 
\beq
\langle T^{a}_{ \ b}\rangle=\frac{K}{12
\pi^{2}}t^{-4}diag(1,-3,1,1)
\eeq
where
\beq
K=\sum_{n=1}^{\infty}\cos(2\pi n 
\alpha)\frac{2+\cosh(na)}{(\cosh(na)-1)^{2}} \label{sum}.
\eeq 
We emphasize that in performing this calculation it is necessary to
justify interchanging the derivative operator in (\ref{unsplit}) with 
the sum in (\ref{Gra}).    This may be justified by observing that any
point $y$ in $IGH$ has a neighbourhood such that, for all pairs 
$(x,x')$ of  points in this neighbourhood, each term in the sum in
(\ref{unsplit}) is differentiable, and moreover, the sums which result
when one acts on each term in (\ref{unsplit}) with one or two unprimed
or primed derivatives are all uniformly convergent.

Sushkov \cite{Sushkov95} pointed out that, since (see \cite{Hiscock82})
the expectation value of the stress-energy tensor for a twisted field is
equal to that for an untwisted field multiplied by a negative constant
(and since a [complex] twisted field may be regarded as an automorphic
field for parameter value $\alpha=1/2$) one expects there to be some
value of $\alpha$ for which the expectation value of the stress-energy
tensor vanishes.  (One can similarly argue that there is another value
of $\alpha$ for which the renormalized expectation value of $\phi^2$
vanishes.)

With the present approach, this argument can be made precise as follows:
If we let the automorphic parameter $\alpha$ equal $1/2$ then $K$ in 
(\ref{sum}) is negative since the first term in the sum (\ref{sum}) is
negative and the terms alternate in sign and decrease monotonically in
absolute value. If we instead let $\alpha$ equal zero then trivially $K$
is positive.  Next, using  the property that $K$ is a continuous
function  of $\alpha$, since every term in the sum is continuous as a 
function of $\alpha$ and since the sum which defines $K$ is uniformly 
convergent in $\alpha$, it immediately follows from the intermediate
value theorem  that there has to exist a value of the automorphic
parameter  such that $K$ and hence the expectation value of the
stress-energy vanishes. 

One way of understanding why the formula (\ref{unsplit}) leads to a 
finite result for the expectation value of $\langle T_{ab}\rangle$ in
the initial globally hyperbolic region is to observe that $G^\alpha_{\rm
ren}$  arises as the difference of two two-point functions, 
$G^{\alpha}$ and  $G_{\rm M}$ each of which have the same local
singularity structure (``Hadamard form'') so that $G^\alpha_{\rm ren}$
itself is  smooth, and, in consequence, the limit in (\ref{unsplit}) is
well-defined (and finite).   ($G_{\rm M}$ plays here the role, which in
the  general procedure (see e.g. \cite{Wald94}) for defining $\langle
T_{ab}\rangle$  is played by the subtraction of a ``locally Hadamard
parametrix''.)   In the setting of a real linear scalar field on a
general spacetime with compactly generated Cauchy horizon, Kay
Radzikowski and Wald \cite{Kay96} showed however (see especially
``Theorem 2 (alternative statement)'' in Section 5 of \cite{Kay96}) that
there are necessarily certain points (``base points'') on the Cauchy
horizon such that in the intersection of any neighbourhood, $N$, of any
base point with the initial globally hyperbolic region, the difference
between any bisolution which takes the Hadamard form on the initial
globally hyperbolic region and any ``Hadamard parametrix'' defined on
$N$ is necessarily singular in the sense that it not only fails to be
smooth,  but fails even to be bounded.  In consequence, $\langle
T_{ab}\rangle$ will necessarily be singular  (or ill-defined) at these
base points as we mentioned in the opening paragraph.  (For exactly
similar reasons, other renormalized quantities such as
$\langle\phi^2\rangle$ will also be singular.)

We now present our direct argument that in the case of complex
automorphic fields on Misner space (where every point on the chronology
horizon is a base point in the sense of \cite{Kay96}) and the Sushkov
state, the statement of the theorem continues to hold.  In other
words:\\ 
{\bf Theorem.} Let $b$ be any point on the chronology horizon of Misner
space, and let  $N$ be any  neighbourhood of $b$. Then, for any value of
the automorphic parameter $\alpha$, there exists at least one pair of
non-null related points $(x,x')\in (N\cap IGH )\times (N\cap IGH)$,
where $IGH$ is the initial globally hyperbolic region,
such that the renormalized two-point function $G^\alpha_{\rm ren}(x, x')$
is singular  (in the sense that $G^\alpha_{\rm ren}(x,y)$ diverges as
$y$ approaches $x'$).\\ 
{\it Proof.} Regarding Misner space as the region $\cal R$ of 
Minkowski space with
the identifications (\ref{ident}), the renormalized two-point function 
$G^\alpha_{\rm ren}$ is given by the formula
(\ref{Gra}) where the square of the geodesic distance is given, in double 
null coordinates by
$$
\sigma(x,L^n x')
=-UV-U'V'+e^{-na}U'V+e^{na}V'U
+(Y-Y')^{2}+(Z-Z')^{2}.
$$
Clearly $N\cap IGH$ will contain a neighbourhood which arises, when we
coordinatize it in this way, as the product of a rectangle $R=\lbrace
(U,V): U_1<U<U_2, V_0<V<0\rbrace$ with a rectangle $S=\lbrace (Y,Z):
Y_1<Y<Y_2, Z_1<Z<Z_2\rbrace$.  So it will suffice to exhibit a pair of
points contained in such a product having the properties stated in the
theorem. In fact, we shall exhibit such a pair for which the $Y$ and $Z$
coordinates are the same -- say $(Y_0,Z_0)\in S$.  Choose a point
$x=(U,V,Y_0,Z_0), (U,V)\in R$.  Then it is easy to see that there exists
a point $\hat x=(\hat U,\hat V,Y_0,Z_0), (\hat U,\hat V)\in R$,
spacelike separated  from $x$, such that every point $x_\delta=(\hat
U,\delta,Y_0,Z_0)$, $\hat V<\delta\le 0$ along  the line connecting
$\hat x$ to the point $(U,0,Y_0,Z_0)$ on the chronology  horizon (see
figure 2) is also spacelike separated from $x$. (Alternatively, one can
choose $\hat x$ to be timelike separated from $x$ and such that all the
points $x_\delta$ are also timelike separated etc.)  We now show that
one can choose $\delta$ so that $L^m x_\delta$ is null related to $x$
for some integer $m$ so that $\sigma(x,L^m x_\delta)=0$. To see this, we
note that for any $\delta$ which takes the form  $$\delta=
\frac{e^{-ma}\hat UV-UV} {\hat U-e^{ma}U}. $$  $L^m x_\delta$ for
$x_\delta=(\hat U,\delta,Y_0,Z_0)$ will be null separated from $x$, and,
by taking $m$ to be sufficiently large and positive, we can arrange for
$\delta$ to lie in the interval $(\hat V,0)$.  Taking  $x'=x_\delta$, we
thus have exhibited a pair $(x,x')$ of points in $(N\cap IGH)\times
(N\cap IGH)$ such that the $m$th term in the sum (\ref{Gra}) is
singular. On the other hand, it is easy to see that the remaining terms
in this sum are uniformly convergent in some neighbourhood of $(x,x')$. 
We conclude that $G_{\rm ren}^\alpha$ is singular at $(x,x')$.$\Box$\\ 

We note that the above theorem can be given a geometrical
interpretation: On the region $\cal R$,  $x_\delta$ has the property
that, in the neighbourhood $N\cap IGH$ it is spacelike separated from
$x$, but one of its images $L^m x_\delta$ is null separated from $x$. 
Reinterpreting $\cal R$ as Misner space by making the identifications
(\ref{ident}), we would rather say that $x$ and $x_\delta$ are spacelike
separated in the neighbourhood $N\cap IGH$ but globally null separated. 
Thus the phenomenon of ``propagation of singularities'' (see the
introductory discussion) is, in the case of  the bisolution $G^\alpha$, 
inherent in the image sum formula.\\\\                                  
\def\epsfsize#1#2{0.5#1}                                                
\epsfclipon                                                             
\epsffile[70 310 775 725]{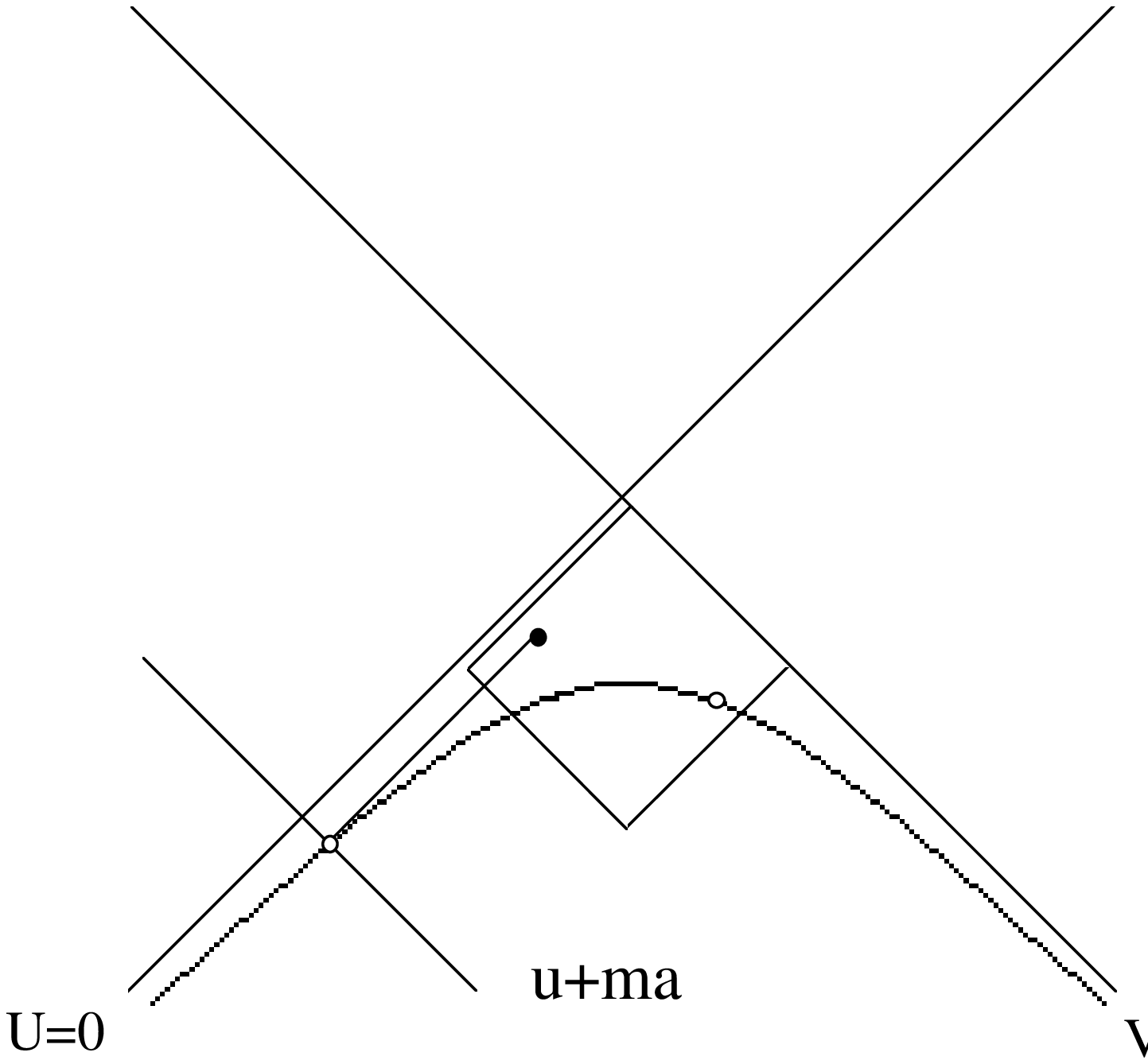}                                    
Fig2.\\\\                                                                
Next, we remark that one might have thought that one could argue that,
since there is a value of $\alpha$ for which $\langle T_{ab}\rangle$
vanishes in the initial globally hyperbolic region, then, by continuity,
for this value of $\alpha$, $\langle T_{ab}\rangle$ would also vanish on
the chronology horizon.   However, the uniform convergence properties
mentioned earlier  do {\it not} hold for neighbourhoods of points {\it
on} the chronology horizon and thus the required continuity property may
(and, in view of the above theorem does!) fail.  What we have exhibited
in fact is a situation where  $\langle T_{ab}\rangle$ has a finite limit
as one approaches the chronology horizon, but nevertheless is singular
on the chronology horizon. Furthermore, and in consequence of this
singularity, we would not regard it as legitimate to view Misner space
together with an automorphic field in the Sushkov state for this value
of $\alpha$ as a solution  to semiclassical gravity.  Thus we would not
regard this model as evidence against the Chronology Protection
Conjecture. 

Finally, we note that one can prove a similar theorem for the analogues
of the Sushkov state on the spacetime (with compactly generated Cauchy
horizon) obtained by taking the product of two-dimensional Misner space
with a two torus.  (In this case one simply adds further terms to the 
image sum (\ref{Gra}), and these terms are non-singular and have a
convergent sum when evaluated at the pair $(x,x')$).  The cases of Gott
space and Grant space are only slightly different.  These have
chronology horizons which are not compactly generated and which contain
no base points (\cite{Kay96}).  Nevertheless the statement of ``Theorem
2'' (but not now of ``Theorem 2 (alternative statement)'' of
\cite{Kay96}) will still hold (see Section 6 of \cite{Kay96}) since
there are null geodesics (on the so-called polarized hypersurfaces --
see \cite{Grant92,Thorne92}) which self-intersect arbitrarily close to
the chronology horizon.  This result can also be reproduced by
elementary methods similar to those of this letter \cite{Cramer-Kay96}
in the case of the states discussed by Boulware \cite{Boulware92} (on
Gott space) and Tanaka-Hiscock \cite{Tanaka96} (on Grant space) (and
also, e.g. the  analogue of the Sushkov state for a  massless
automorphic field on  Gott space or Grant space).  In fact, all these
states are  singular in the neighbourhood of any point on the chronology
horizon in the sense that a statement similar to the above theorem still
holds, where, however, the pair $(x,x')$ must now be chosen so that at
least one of $x$, $x'$ lies in the ``CTC'' side of the chronology
horizon (where one extends the two-point functions of the states to this
side of the horizon in an obvious way).  
\vskip 1 cm 
\noindent 
We wish to thank Chris Fewster, Atsushi Higuchi, Ted Jacobson, and 
James Vickers for valuable  discussions.   

\end{document}